\documentclass[pre,twocolumn,superscriptaddress,showpacs]{revtex4-1}

\usepackage{graphicx}
\usepackage{amsmath}
\usepackage{latexsym}
\usepackage{comment}
\usepackage{bm}

\begin{document}
\title{Theory of cylindrical dense packings of disks}
\date{\today}
\author{A. Mughal}
\affiliation{Institute of Mathematics and Physics, Aberystwyth University, Penglais, Aberystwyth, Ceredigion, Wales, UK,SY23 3BZ}
\affiliation{Theoretische Physik, Fried.-Alex.-Universit\"{a}t Erlangen-N\"{u}rnberg - Staudtstr. 7, 91058 Erlangen, Germany, EU}
\author{D. Weaire} 
\affiliation{Foams and Complex Systems, School of Physics, Trinity College Dublin, Dublin 2, Ireland}

\begin{abstract}
We have previously explored cylindrical packings of disks and their relation to sphere packings \cite{Mughal:2011} \cite{Mughal:2012} \cite{Mughal:2013}. Here we extend the analytical treatment of disk packings, analysing the rules for phyllotactic indices of related structures and the variation of the density for line-slip structures, close to the symmetric ones. We show that rhombic structures, which are of a lower density, are always unstable i.e. can be increased in density by small perturbations
\end{abstract}

\maketitle

% -----------------------------------------------------------------------------
\section{Introduction}

In a previous papers (see \cite{Mughal:2011}, \cite{Mughal:2012}, \cite{Mughal:2013}, \cite{Pickett:2000},  \cite{chan:2013} and \cite{chan2011densest}) simulation techniques have been applied to the packing of hard spheres within a cylinder, together with an analysis of the related problem of packing disks on the surface of a cylinder. 

Many distinct sphere packings were identified, as the ratio $D/d$ of the diameters of the cylinder and sphere was varied up to about $D/d=2.873$. For $D/d$ below 2.71486, the densest structures consisted entirely of spheres in contact with the cylindrical wall.

Up to that point, the disk packings on the surface were of a similar character, and these could be described analytically. An approximate analytic correspondence to the sphere packings was established, and hence their nature and sequence of the sphere packings could be interpreted, semi-quantitatively.

Apart from special cases at very low $D/d$, all of these structures are of the same character. For certain discrete values of $D/d$, a close-packed phyllotactic structure was found. Between these values, the misfit with the cylinder surface was taken up by the incorporation of a so-called line-slip, in which adjacent parallel lines of spheres were displaced to form a spiral line defect in an otherwise close-packed structure.

Finer details included a square-root singularity at certain points in the variation of the packing fraction with $D/d$, and various structures of slightly lower packing fraction (other line-slips, and affine sheared structures).

In the present paper we amplify and extend the theoretical analysis in various directions. These include consideration of the competition between alternative line-slips (since three possibilities present themselves in each case) and the stability of structures. 

The main body of the paper is devoted to the disk packing problem, for which our previous analytical treatment can be extended. 

\begin{figure}
\begin{center}
\includegraphics[width=0.9\columnwidth ]{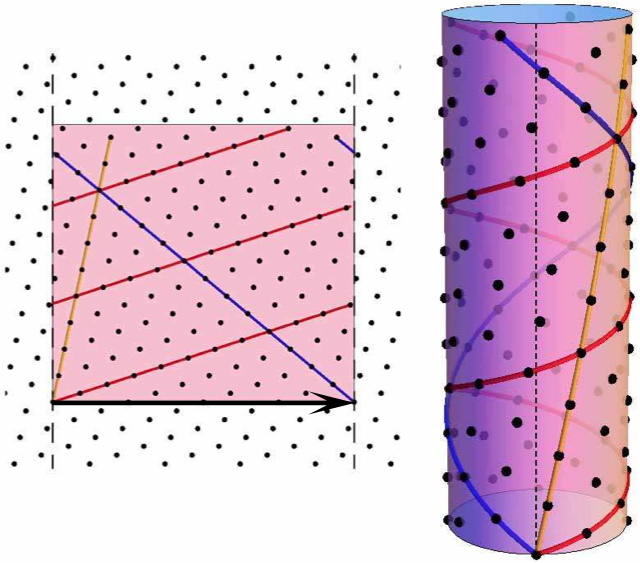}
\caption{Left: a triangular lattice with a periodicity vector vector ${\bf V}$ between two lattice points, shown by a black arrow. Right: The excised section wrapped onto a cylinder of diameter $|{\bf V}|/\pi$.}
\label{cwrap}
\end{center}
\end{figure}

\section{Phyllotactic Notation}

In spite (or perhaps  because) of the antiquity of the study of phyllotaxis, its formal expression is often unclear, so we will summarise it here in terms appropriate to the present work.

We begin by considering the problem of seamlessly wrapping a symmetric triangular lattice onto a cylinder as shown in Fig \ref{cwrap}. We take the lattice spacing (nearest neighbour distance) to be unity. The seamless wrapping that we seek is possible only if we can define a periodicity vector ${\bf V}$ between a pair of lattice points, shown by the black arrow on the left hand side of Fig (\ref{cwrap}), which is commensurate with the diameter of the cylinder as follows. We can define two edges, at the base and the head of ${\bf V}$, both of which are perpendicular to ${\bf V}$. After cutting along the edges the excised section can be wrapped around a cylinder of diameter $|{\bf V}|/\pi$, as shown on the right hand side of Fig \ref{cwrap} (where the cut edges meet along the dashed line on the cylinder).

%put V on arrow

A cylindrical pattern created in this way consists, in general, of spiral lines in three directions; exceptional cases include the limiting case of lines that  go around the circumference or are parallel to the cylinder axis. On the plane the spiral lines, shown by the red, blue and yellow lines in Fig \ref{cwrap}, correspond to rays traced out along the direction of the primitive lattice vectors. Each of these three directions may be associated with a {\it phyllotactic index} (for which the traditional term in biology is {\it parastichy}). This is a positive integer which answers the question: how many such spirals do I need before the cylindrical pattern is complete? %perhaps "repeats" would be better

%how many times does the spiral line wrap around the cylinder before the pattern is complete. %confused!

\begin{figure}
\begin{center}
\includegraphics[width=1.0\columnwidth ]{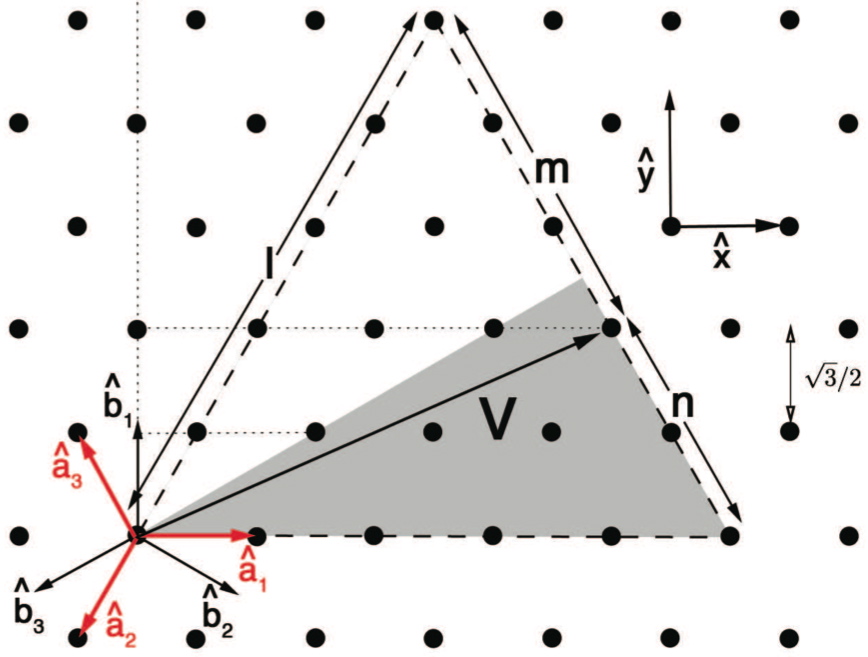}
\caption{A symmetric triangular lattice on to which is inscribed the periodicity vector ${\bf V}$ with phyllotactic indices [l=5,m=3,n=2]. By convention the periodicity vector is confined to the shaded region (see text). The diagram shows the Cartesian unit vectors ${\bf {\hat x}}$ and ${\bf {\hat y}}$. The unit primitive lattice vectors of the triangular lattice (${\bf a_i}$) and the unit basis vectors (${\bf b_i}$) used in the text are shown by small red and black arrows, respectively. Also shown is an equilateral triangle: the phyllotactic indices are related to the length of the triangle sides and segments thereof, as indicated.}
\label{phyllotactic_indices}
\end{center}
\end{figure}

In this way any such periodicity vector ${\bf V}$ van be assigned a unique set of three positive integers as indices $[l,m,n]$, where $m\geq n$ and $l=m+n$. There are a number of ways to see this. 

The first is a working definition in terms of a diagram. Consider as an example the periodicity vector shown in Fig (\ref{phyllotactic_indices}). Also shown is a an equilateral triangle, such that the base of ${\bf V}$ coincides with one of the corners of the triangle while the head of ${\bf V}$ is located on the opposing side. We adopt the convention that ${\bf V}$ always lies within the shaded area, without loss of generality, when symmetry is taken into consideration. The triangle has sides of length $l$ and the head of the vector subdivides opposing side into segments of length $m$ and $n$. Thus, the there phyllotactic indices are related to the length of the triangle sides and segments thereof, as indicated.

Another way to understand the assignment of phyllotactic indices is by noting that the lattice rows in a particular direction divide the plane into strips of width $\sqrt{3}/2$; there are two such strips parallel to the unit vector ${\bf {\hat a}_1}$ which cross ${\bf V}$ (as indicated in Fig (\ref{phyllotactic_indices})), and this means that one of the indices takes the value 2. By considering the strips crossing ${\bf V}$ parallel to ${\bf {\hat a}_2}$ and ${\bf {\hat a}_3}$ it can be seen that the other two phyllotactic indices are 3 and 5, respectively.

A more direct formal description follows with definitions that we will use later. The nearest neighbour vectors are 
\begin{eqnarray}
{\bf {\hat a _1}}&=&(1,0)
\nonumber
\\
{\bf {\hat a _2}}&=&(-1/2, \sqrt{3}/2)
\nonumber
\\
{\bf {\hat a _3}}&=&(-1/2, -\sqrt{3}/2).
\nonumber
\end{eqnarray}
A second set of unit vectors can be obtained by rotating the unit primitive lattice vectors by $\pi/2$ (as shown in Fig (\ref{phyllotactic_indices})), giving
\begin{eqnarray}
{\bf {\hat b _1}}&=&(0,1)
\nonumber
\\
{\bf {\hat b _2}}&=&(\sqrt{3}/2, -1/2)
\nonumber
\\
{\bf {\hat b _3}}&=&(-\sqrt{3}/2, -1/2)
\nonumber
\end{eqnarray}
where $\sum_{i=1}^3 \bf {\hat b}_i=0$. The projection of the periodicity vector onto the $i$th vector is given by $V_{i}= {\bf V} \cdot {\bf b}_i$. Any lattice vector ${\bf V}$ may be associated with phyllotactic indices $l$,$m$, and $n$ which are the ordered (decreasing) absolute values of
\begin{equation}
c_{i}=\frac{V_i}{\sqrt{3}/2},
\label{eq:coef}
\end{equation}
where $\sum_i^3 c_i=0$. Specifically $c_1=n$, $c_2=l-n=m$ and $c_3=-l$. The indices $c_i$ will be useful in the analysis which follows in later sections.

\section{Rhombic structures and their notation}

\begin{figure*}
\begin{center}
\includegraphics[width=2.0\columnwidth ]{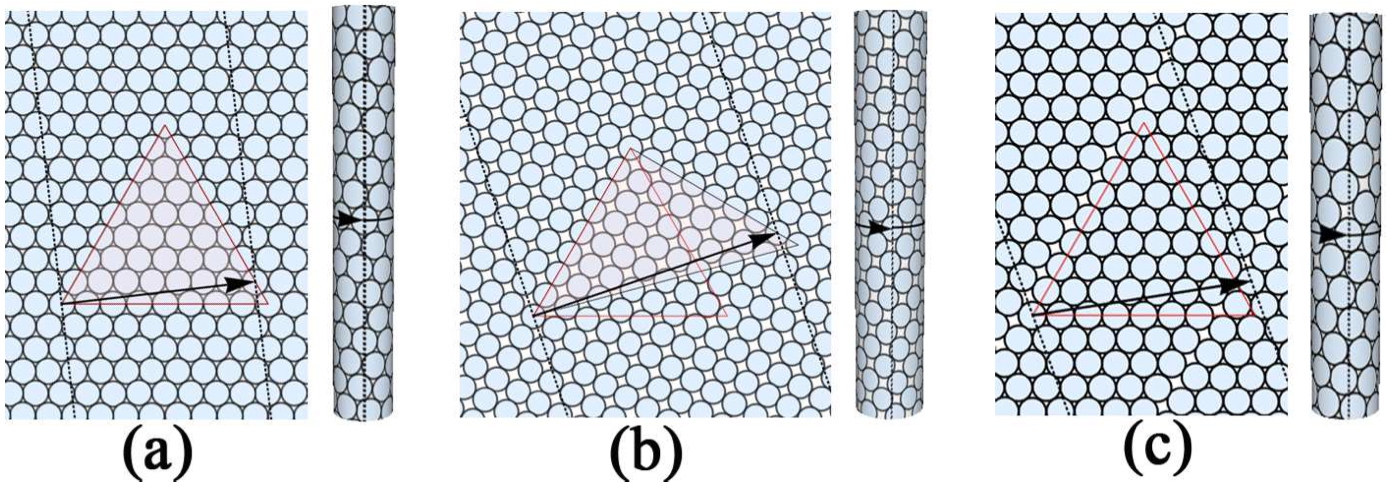}
\caption{The three disk packings discussed here are: {\bf (a)} symmetric packings (where the disks are centred on a triangular lattice - see Fig (\ref{disk_packings}a), (b) affine structures and (c ) line-slip packings. The black arrow is the periodicity vector ${\bf V}$. The region between the dashed lines can be excised and wrapped on to a cylinder of the appropriate diameter.}
\label{disk_packings}
\end{center}
\end{figure*}

\begin{figure}
\begin{center}
\includegraphics[width=0.8\columnwidth ]{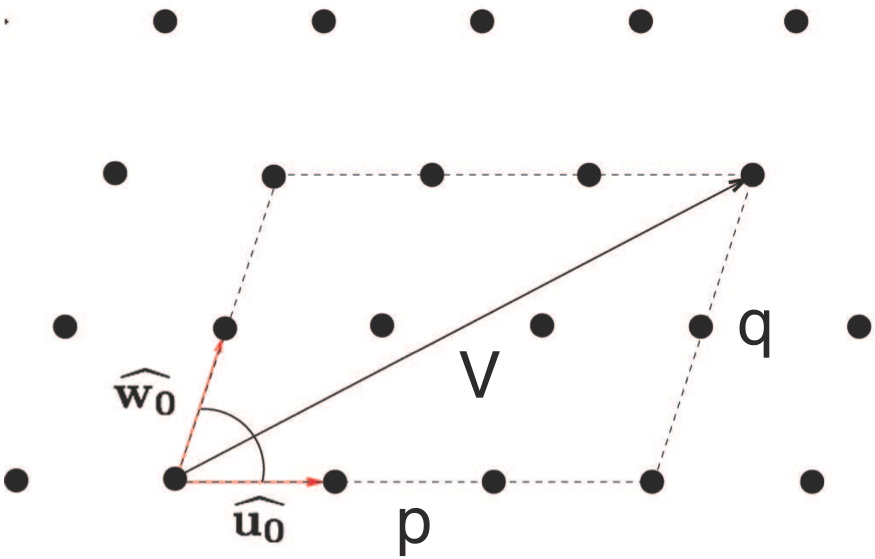}
\caption{A rhombic lattice onto which is inscribed the periodicity vector ${\bf V}$ with phyllotactic indices $[p=3, q=2]$. The primitive lattice vectors ${\bf {\hat u}_0}=(1,0)$ and ${\bf {\hat v}_0}=(\cos(\theta),\sin(\theta))$ are shown by the red arrows.}
\label{rhombic}
\end{center}
\end{figure}

As explained above, a 2D triangular close-packed arrangement of disks on a plane can be wrapped onto the surface of a cylinder of an appropriate diameter, as in Fig (\ref{disk_packings}a). The most obvious way in which this structure can be adjusted to be consistent with an arbitrary diameter (that is, to have a vector ${\bf V}$ of the corresponding magnitude) is by an affine deformation. An appropriate affine deformation can create a structure in which all of the contacts in one direction become separated, while the others are maintained, see Fig (\ref{disk_packings}b). 

This may be called a rhombic structure, since the contact vectors form a rhombus, as in Fig (\ref{rhombic}). When such structures were investigated in the 3D packings, they were found to be always of a lesser density than line-slip structures, and they were thought to be unstable with respect to small perturbations. Here we will analyse the question of stability, but only for the corresponding disk packing problem in 2D. We find that the rhombic packings are indeed always unstable. 

\begin{figure}
\begin{center}
\includegraphics[width=1.0\columnwidth ]{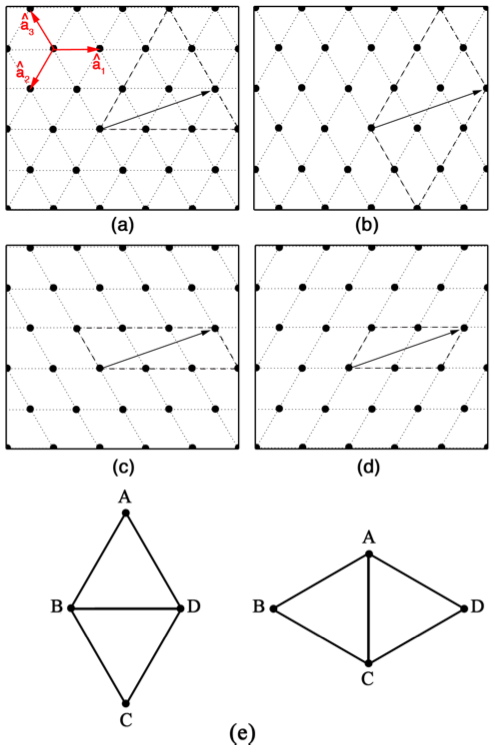}
\caption{(a) a triangular lattice onto which is inscribed a periodicity vector ${\bf V}$ with phyllotactic indices $[l,m,n]$. A rhombic lattice can be formed by disregarding the lattice lines in the (b) $\widehat{\bf a_1}$ direction, (c) $\widehat{\bf a_2}$ direction or (d) $\widehat{\bf a_3}$ direction. The rules that connect the two sets of indices are given in the text. (e) If contact BD is broken and BD is lengthened, a new contact AC is eventually established.}
\label{triangular_rhombus}
\end{center}
\end{figure}

Consider a rhombic lattice as shown in Fig (\ref{rhombic}). It is a lattice in which the fundamental region is a rhombus of side length unity and has angles $\theta$ (here called the rhombic angle) and $\pi-\theta$, where $\theta=\pi/3$ and $\theta=2\pi/3$ correspond to a triangular lattice while $\theta=\pi/2$ yields a square lattice. Such a rhombic lattice can be seamlessly wrapped onto a cylindrical surface as explained above for the triangular lattice, with the choice of $\theta$ providing the flexibility to adjust to a particular cylinder diameter.

For the rhombic lattice there are two unit primitive lattice vectors ${\bf {\hat u}_0}$ and ${\bf {\hat w}_0}$, and any periodicity vector can be written in terms of these as,
\[
{\bf V}=p{\bf {\hat u}_0} + q{\bf {\hat w}_0} 
\]
where the indices $p$ and $q$ are positive (with $p\geq q$) if ${\bf {\hat u}_0}$ and ${\bf {\hat w}_0}$ are suitably chosen. These are the phyllotactic indices for the rhombic lattice. Again they count the number of lattice strips that cross the periodicity vector ${\bf V}$.

Of course, the triangular lattice of Fig (\ref{phyllotactic_indices}) may be turned into a rhombic one by disregarding the lines in any of the three directions, as in Fig (\ref{triangular_rhombus}). Imposing the convention described in Fig (\ref{phyllotactic_indices}), there is an obvious relation between the two kinds of indices. Consider a triangular lattice onto which is inscribed a periodicity vector {\bf V} with the phyllotactic indices $[l,m,n]$. By disregarding the lattice lines in the ${\bf {\hat a}_1}$, ${\bf {\hat a}_2}$ and ${\bf {\hat a}_3}$ directions - as shown in Fig (\ref{triangular_rhombus}b-\ref{triangular_rhombus}d)  - we deduce the following rules between the two sets of indices, appropriately defined,
\begin{eqnarray}
\textrm{disregarding}\;\;
{\bf {\hat a}_1}
&:&
\;\;\;
[p=l, q=m]
\nonumber
\\
\textrm{disregarding}\;\;
{\bf {\hat a}_2}
&:&
\;\;\;
[p=l, q=n]
\nonumber
\\
\textrm{disregarding}\;\;
{\bf {\hat a}_3}
&:&
\;\;\;
[p=m, q=n].
\nonumber
\end{eqnarray}

Beginning with a triangular lattice, the affine deformation preserves the rhombic symmetry with the loss of one nearest-neighbour contact - let it be contact $i$ in direction $i$. The ``strip-counting'' identification of the phyllotactic indices clearly shows that the index corresponding to the strips in direction $i$ is lost, when the other two remain as the rhombic indices, as above, the question is: when triangular symmetry is restored by taking the deformation to its limit, and making new contacts as indicated by Fig (\ref{triangular_rhombus}e), what is the new ordered set of indices $l',m',n'$? 

We distinguish various cases, as follows immediately.
\underline{Case 1: $(l,m,n)=(l,l,0)$} In this case ${\bf V}$ points in one of the nearest neighbour directions, corresponding to the third index. If one of the other directions is chosen to break a contact then $|{\bf V}|$ is unchanged, hence the indices of ${\bf V}$ must be unchanged and
\[
(l',m',n')=(l,m,n)=(l,l,0)
\]
The invariance of ${\bf V}$ is only possible in this case. If instead the contact in the direction of ${\bf V}$ is broken, ${\bf V}$ is increased and the only possibility is
\[
(l',m',n')=(2l,l,l)
\]

\underline{Case 2:} The remaining case is specified by $(l,m,n)$ with $l>m$, hence $n>0$. Preservation of all three indices is impossible (see above). It follows from the various inequalities that the only logical possibilities are:

(i) Preserve $l$ and $m$
\[
(l',m',n')=(l+m,l,m)
\]

(ii) Preserve $l$ and $n$
\[
(l',m',n')=(l+n,l,n)
\]

(iii) Preserve $m$ and $n$
\[
(l',m',n')
 =
\left\{
\begin{array}{l l}
(m,n,m-n) & \quad \mbox{if $n\geq m-n$}\\
\;\;\;\;\;\;\;\; \\
(m,m-n, n) & \quad \mbox{if $n\leq m-n$} \\
\end{array}
\right.
\]

\section{Stability}

\begin{figure*}
\begin{center}
\includegraphics[width=2.0\columnwidth ]{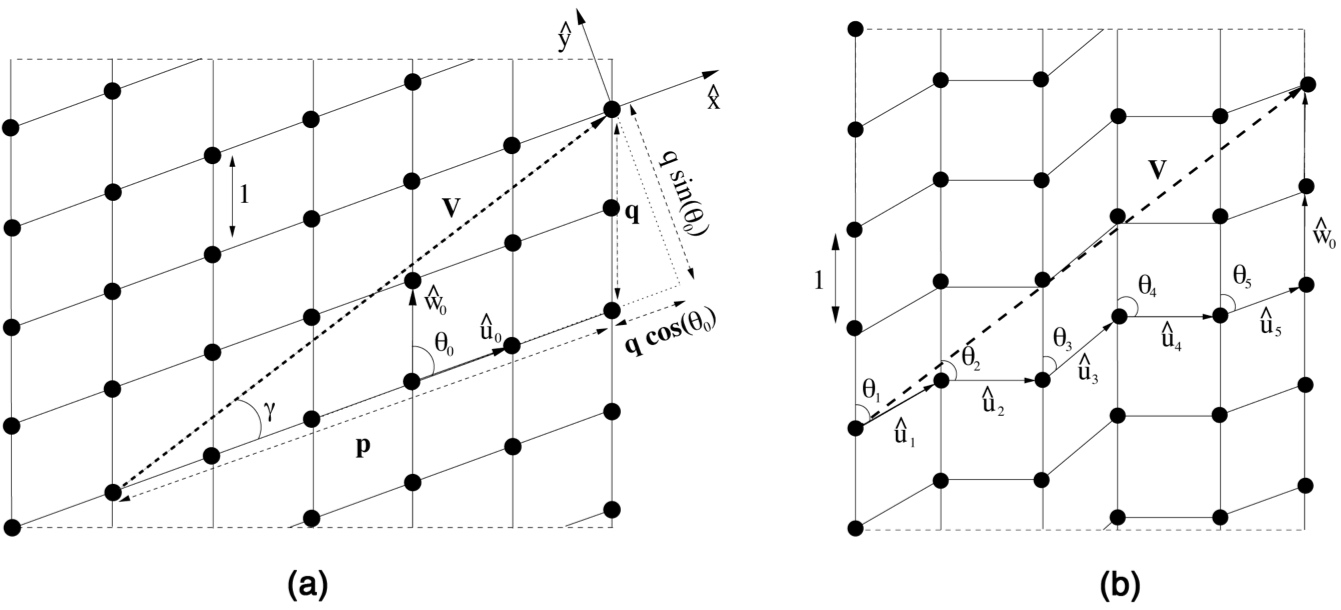}
\caption{ {\bf (a)} An unperturbed rhombic lattice (with sides of length 1) onto which is inscribed a periodicity vector ${\bf V_0}$ with phyllotactic indices $[p,q]$. The angle between the rhombic lattice vectors ${\bf u}_0$ and ${\bf w}_0$ is given by $\theta_0$; the angle between the $\bf V$ and ${\bf u}_0$ is denoted by $\gamma$. {\bf (b)} A perturbation of the rhombic lattice. The lattice is divided into $p$ strips and within each strip the rhombic angle is deviate from $\theta_0$. In fact such a perturbation would involve a rotation of the resulting structure which, for convenience, is not shown in the image.} 
 \label{unbuckled}
\end{center}
\end{figure*}

In this section we investigate the stability of the rhombic lattice.

We begin by considering a lattice with a rhombic angle $\theta=\theta_0$, as shown in Fig (\ref{unbuckled}). Note that we have set the unit vector ${\bf {\hat u}_0}$ along the x-axis, so that the components of ${\bf V_0}$ are given by
\begin{equation}
V_{0x}=p+q\cos(\theta_0) \;\;\; \textrm{and} \;\;\; V_{0y}=q\sin(\theta_0).
\label{eq:components}
\end{equation}
and $\tan({\gamma})=V_{0y}/V_{0x}$ where the angle $\gamma$ is as indicated in Fig (\ref{unbuckled}).

We now consider a  perturbation of the rhombic lattice along the $\widehat{\bf u}_0$ direction as shown in Fig (\ref{unbuckled}a). The lattice is divided into $p$ parallel strips (labelled $1,2,3,4$...$p$), which cross ${\bf V}$. Where the rhombi in the $i$th strip have a rhombic angle
\begin{equation}
\theta_i=\theta_0+\Delta\theta_i,
\label{eq:perturbation}
\end{equation}
where $\theta_0$ is the rhombic angle in the absence of any perturbation and is restricted to lie between $\pi/3$ and $2\pi/3$; note for these two values the unperturbed lattice has triangular symmetry with an additional contact. 

Alternatively, a similar perturbation could be been imposed in the $\widehat{\bf w}_0$. This would involve decomposing the lattice into $q$ parallel strips that cross ${\bf V}$. However, as we shall demonstrate, this mode is always unstable and therefore not of further interest.

As shown in Fig (\ref{unbuckled}b), the perturbation to the rhombic angle of the $i$th strip is $\Delta \theta_i$ which we normalise as follows,
\[
\Delta \theta_i=\lambda \widehat{\Delta \theta_i},
\]
where $\widehat{\Delta \theta_i}$ are the components of a unit vector so that %remove unit vector
\begin{equation}
\sum_i^p \widehat{\Delta \theta_i}^2=1,
\label{eq:rhombic_ident_1}
\end{equation}
and for convenience in what follows we also define
\begin{equation}
\sum_i^p \widehat{\Delta \theta_i}=\mu.
\label{eq:rhombic_ident_2}
\end{equation}
It will turn out that relevant quantities depend only on $\lambda$ (the strength of the perturbation) and $\mu$. So we will be writing equations for many perturbations which have the same values of these parameters. Note that we have not yet imposed the condition that $|{\bf V|}$ is constant%need to edit this.

For the unperturbed lattice the area of the rhombus is, 
\[
A=\frac{1}{p}\sum_i^p\sin(\theta_0),
\]
while in the perturbed case the {\it average} area is,
\begin{eqnarray}
A'
&=&
\frac{1}{p}\sum_i^p\sin(\theta_0+\lambda \widehat{\Delta \theta_i})
\\
&=&
\frac{1}{p}\sum_i^p
\sin(\theta_0) + \lambda \widehat{\Delta \theta_i} \cos(\theta_0)-  \frac{1}{2}\left(\lambda \widehat{\Delta \theta_i}\right)^2 \sin(\theta_0).  
\nonumber
\end{eqnarray}
where we have expanded to second order in $\lambda$. The change in the average area is,
\begin{equation}
\delta A=A'-A
=
\frac{1}{p}
\left(
\mu\lambda\cos(\theta_0)-\frac{\lambda^2}{2}\sin(\theta_0)
\right)
\label{eq:deltaA}
\end{equation}
where we have used Eq. ($\!\!$~\ref{eq:rhombic_ident_1}) and Eq. ($\!\!$~\ref{eq:rhombic_ident_2}).

We now estimate the corresponding change in the length of ${\bf V}$, which we will require to be zero. In the case of the unperturbed lattice we can write the periodicity vector in terms of two primitive unit lattice vectors giving ${\bf V}=p\widehat{\bf u}_0 + q\widehat{\bf w}_0$, however, after perturbation we have, as illustrated in  Fig (\ref{unbuckled}b),
%\[
%{\bf V}=\sum_{i}^N \widehat{\bf u}_i + L \widehat{\bf w}_0. 
%\]
\[
{\bf V}=  \sum_{i}^p \widehat{\bf u}_i + q \widehat{\bf w}_0
\]
It is clear from Eq. ($\!\!$~\ref{eq:perturbation}) that this local perturbation corresponds to a rotation for each vector $\widehat{\bf u}_i$. We can estimate components of the vectors $\widehat{\bf u}_i$, to leading order, as follows. Let a unit vector $\widehat{\bf u}_i$ be initially aligned parallel to the $x$-axis so that it has components $\widehat{\bf u}_i=(1,0)$. After a small rotation by an angle $\Delta \theta_i$ its new components, to leading order, are $\widehat{\bf u'}_i=(1-(\Delta \theta_i)^2/2,\Delta \theta_i)$. Thus the change in the periodicity vector is,
\begin{eqnarray}
\delta {\bf V}
&=&
\sum_{i}^p \delta{\bf u}_i=\sum_{i}^p (\widehat{\bf u'}_i - \widehat{\bf u}_i)
\nonumber
\\
&=&
\sum_i^p \left(  -\frac{\lambda^2 \widehat{\Delta \theta_i}^2 }{2}, \lambda \widehat{\Delta \theta} \right)
\nonumber
\\
&=&
(-\lambda^2/2, \lambda\mu)
\label{eq:deltaV}
\end{eqnarray}
To leading order the length of the periodicity vector, after perturbation, is then,
\begin{eqnarray}
|{\bf V}|^2
&=&
({\bf V_0}+\delta{\bf V})^2
\nonumber
\\
&=&
V_0^2 + 2\delta{\bf V}\cdot {\bf V_0} + \delta V^2
\label{eq:Vsquare}
\end{eqnarray}
where $ {\bf V_0}= (V_{0x},  V_{0y})$. In considering the stability of an affine structure we impose the constraint that the length of the periodicity vector remains constant, that is we require $|{\bf V}|^2-V_0^2=0$, which gives the condition,
\begin{equation}
-V_{0x}+2\frac{\mu}{\lambda} V_{0y}+\frac{\lambda^2}{4}+\mu^2=0,
\label{eq:Vcondition}
\end{equation}
to second order in $\lambda$. Rearranging Eq. ($\!\!$~\ref{eq:Vcondition}) gives,
\[
\mu=\frac{\lambda V_{0x}}{2V_{0y}}-\frac{\lambda}{2V_{0y}}\left( \frac{\lambda^2}{4} + \mu^2\right)
\]
by substituting this expression $\mu$ into the right hand side it is possible to recursively developed an ascending power series in $\lambda$. In the limit $\lambda \rightarrow 0$ this gives
\begin{equation}
\frac{\mu}{\lambda}=\frac{V_{0x}}{2V_{0y}}.
\label{eq:mu}
\end{equation}
Substituting  Eq. ($\!\!$~\ref{eq:mu}) into Eq. ($\!\!$~\ref{eq:deltaA}) we have the required expression for the change in area as a function of $\lambda$,
\begin{equation}
\delta A=\frac{\lambda^2}{2p}\sin(\theta_0)\left( \cot(\gamma) \cot(\theta_0) -1\right),
\label{eq:delta_A}
\end{equation}
where  $\cot(\gamma)=V_{0x}/V_{0y}$. 

\subsection*{Condition for stability for displacements along the $\underline{\widehat{\bf u}_0}$ direction}

\begin{figure}
\begin{center}
\includegraphics[width=1.0\columnwidth ]{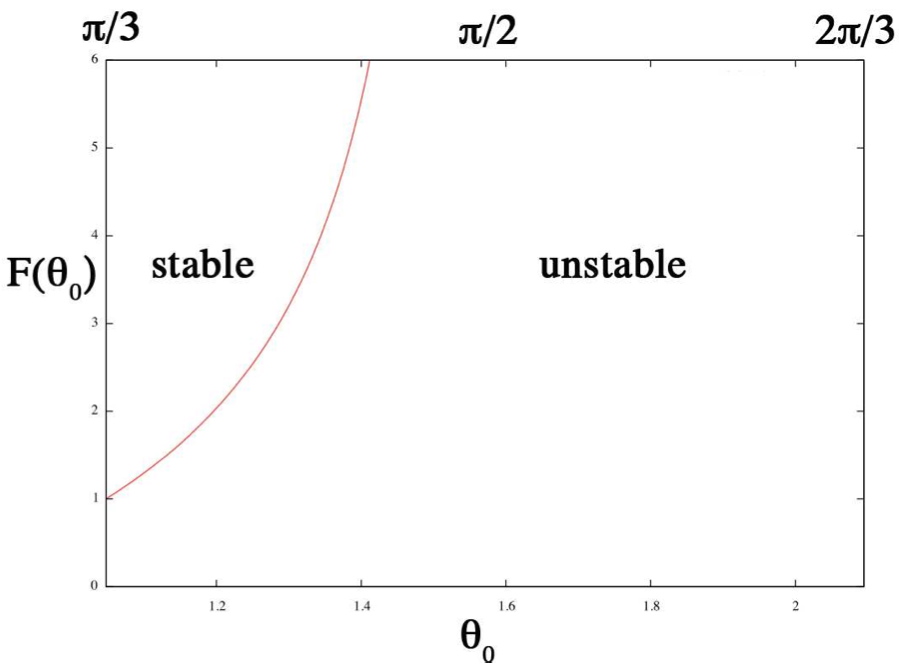}
\caption{A plot of $F(\theta_0)$ over the range $\pi/3 \leq \theta \leq 2\pi/3$ - which determines stability according to  Eq. ($\!\!$~\ref{eq:stabilityzzz}) }
\label{stability_graph}
\end{center}
\end{figure}

In order for rhombic lattice to be stable we require that $\delta A>0$. Thus, from Eq. ($\!\!$~\ref{eq:delta_A}) we have,
\[
\frac{\lambda^2}{2p}\sin(\theta_0)\left( \cot(\gamma) \cot(\theta_0) -1\right)>0,
\]
which reduces to,
\[
 \cot(\gamma) \cot(\theta_0) =1
\]
or more simply
\begin{equation}
\frac{V_{0x}}{V_{0y}} > \tan(\theta_0).
\label{eq:ratio_condition}
\end{equation}
Using Eq. ($\!\!$~\ref{eq:components}) we can write Eq. ($\!\!$~\ref{eq:ratio_condition}) as the condition for stability.
\begin{equation}
\frac{p+q\cos(\theta_0)}{q\sin(\theta_0)} > \tan(\theta_0)
\label{eq:stabilityxxx}
\end{equation}
which can be rearranged to give
\[
p+q\cos(\theta_0)> \frac{q\sin^2{\theta_0}}{\cos(\theta_0)}
\]
or,
\[
\frac{p}{q}>\frac{\sin^2(\theta_0)-\cos^2(\theta_0)}{cos(\theta_0)}
\]
which can be written as
\begin{equation}
\frac{p}{q}> - \frac{\cos(2\theta_0)}{\cos(\theta_0)}=F(\theta_0).
\label{eq:stabilityzzz}
\end{equation}
plotting $F(\theta_0)$, see Fig (\ref{stability_graph}), we find that depending on the ratio $p/q$ the rhombic lattice is either stable or unstable as a function of the rhombic angle $\theta_0$.

\subsection*{Condition for stability for displacement along the $\underline{\widehat{\bf w}_0}$ direction}

We now turn to the alternative case where the perturbation is along the $\widehat{\bf w}_0$ direction and show that it is always unstable. In this case the condition for stability follows exactly as above except that the role of the indices $p$ and $q$ are interchanged, so that  Eq. ($\!\!$~\ref{eq:stabilityxxx}) becomes 
\begin{equation}
\frac{q+p\cos(\theta_0)}{p\sin(\theta_0)} > \tan(\theta_0).
\label{eq:stabilityyyy}
\end{equation}
Simplifying Eq. ($\!\!$~\ref{eq:stabilityxxx}) gives the condition
\[
\frac{q}{p}> - \frac{\cos(2\theta_0)}{\cos(\theta_0)}=F(\theta_0).
\]

\subsection*{Conclusion regarding stability}

The combination of the two above conditions yields the result that a rhombic lattice is always unstable with respect to at least one of the distortions considered.

\section{Line-slip structures}

\begin{figure*}
\begin{center}
\includegraphics[width=2.0\columnwidth, ]{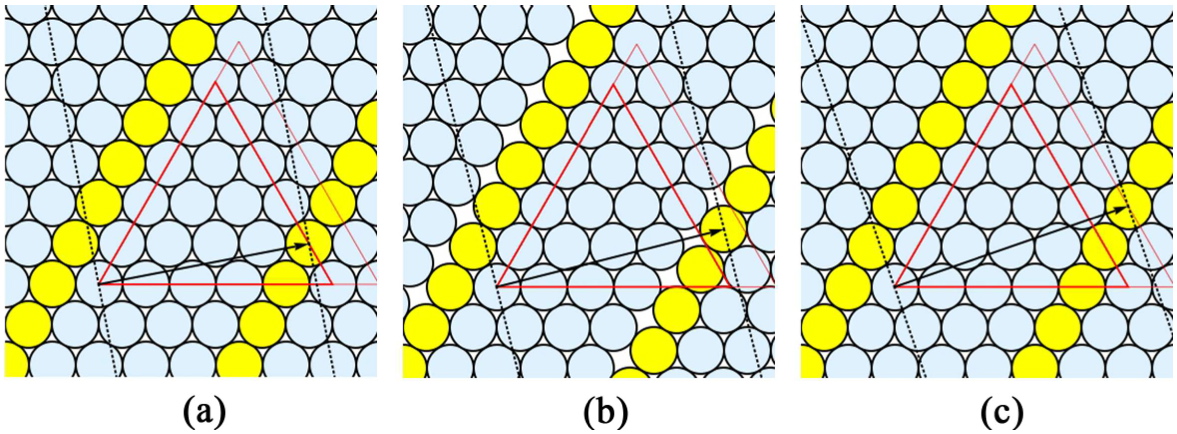}
\caption{({\bf a}) Symmetric packing [5,4,1], ({\bf b}) line-slip in the ${\bf -a_2}$ direction and ({\bf c}) symmetric packing [6,4,2]. Note that for the line-slip the blue disks remain fixed while the yellow disks are displaced.}
\label{lines_slip_highlighted}
\end{center}
\end{figure*}

As noted previously, for specific values of the periodicity vector ${\bf V}$ it is possible to wrap a symmetric disk packing onto the surface of a cylinder. However, around these special points there exist variety of line-slip structures (see Fig (\ref{disk_packings}c) which have packing fractions that vary continuously (and usually linearly) from the value of the symmetric arrangement. Here we will analyse these variations close to the symmetric structures.

Around each symmetric arrangement there are a total of six line slips, two for each of the primitive lattice vectors of the triangular lattice. Let us describe one of these line-slip arrangements with reference to the symmetric packing [5,4,1].  Fig (\ref{lines_slip_highlighted}) shows the symmetric arrangement $[5,4,1]$, we allow the highlighted lattice row to slide in the $-{\bf a_2}$ direction. Ultimately this continues until the symmetric arrangement $[6,4,2]$ is reached. Note that in this case four lattice rows remain fixed while one is allowed to vary. Similarly instead the slip could have been in the ${\bf a_2}$ direction (i.e. the opposite direction) which would have lead to the state $[4,4,0]$. There are two other lattice directions and for each there are two possibilities (and in each case the number of rows which remain fixed depends on the direction of the line-slip). This gives the total of six.

We will analyse the variation of density $\Pi$ with $|{\bf V}|$ for line-slip structures close to the symmetric ones. We will derive the first and second order derivatives at these points which may be used in an expansion of the form,
\begin{equation}
{\Pi}=\Pi_0 
+ 
\left. \frac{d\Pi}{dV} \right|_{V=V_0} \delta V
+ 
\left. \frac{1}{2}\frac{d^2\Pi}{dV^2}\right|_{V=V_0} (\delta V)^2 
\label{eq:density_expansion}
\end{equation}
This serves to elucidate many of the qualitative features of the numerical results, i.e. degeneracies and the choice of the line-slip which gives the maximum density.

\begin{figure*}
\begin{center}
\includegraphics[width=2.0\columnwidth ]{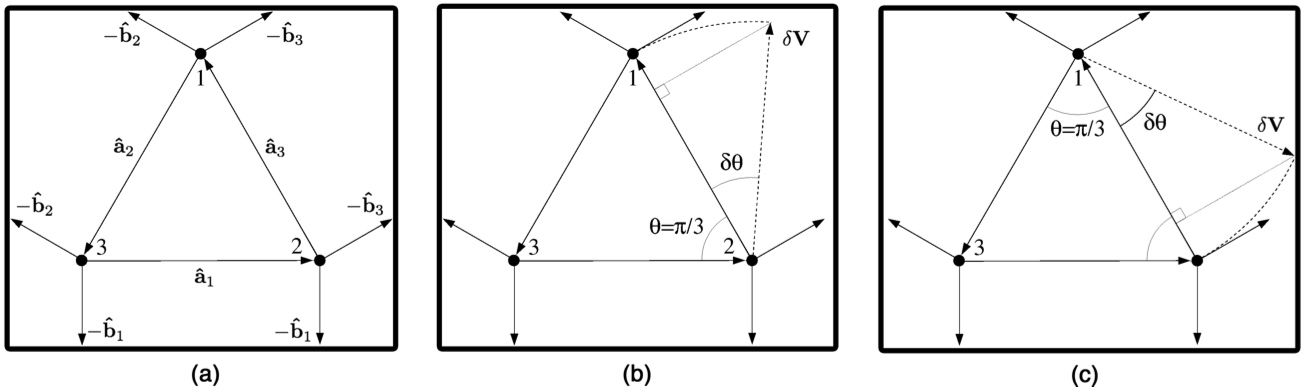}
\caption{Diagram used to analyse the effect of a line-slip. A change in the vertex angle by $\delta\theta$ leads to a variation in the length of the periodicity vector ${\bf V}$ and the density.}
\label{fundamental_triangle}
\end{center}
\end{figure*}

We begin with the construction shown in Fig (\ref{fundamental_triangle}a). This shows an equilateral triangle whose sides can be traversed in a anti-clockwise direction by following the unit primitive vectors ${\bf a}_1$, ${\bf a}_2$ and ${\bf a}_3$. Assume that the periodicity vector ${\bf V}$ terminates on one of the vertices labelled $1$, $2$ or $3$. 

For purposes of demonstration let us assume that the periodicity vector terminates on vertex $1$. Then as shown in Fig (\ref{fundamental_triangle}b), ${\bf V}$ can be perturbed by a rotation by an angle $\delta \theta$ about axis centred on vertex $2$ (i.e. a rotation about tail of the vector ${\bf a}_3$) . Let us denote the change in ${\bf V}$ by the vector ${\bf \delta V}$. This can be decomposed into a component parallel to ${\bf a}_3$ and (perpendicular to this) is a component parallel to ${\bf b}_3$, we have
\[
{\bf \delta V}
=
-\sin(\delta\theta)
{\bf b}_3
+
(1-\cos(\delta\theta))
{\bf a}_3.
\]
Alternatively if the rotation is centred on vertex $1$ (i.e. a rotation about the head of the vector ${\bf a}_3$), as shown in Fig (\ref{fundamental_triangle}c), the change in ${\bf V}$ is given by, 
\[
{\bf \delta V}
=
-\sin(\delta\theta)
{\bf b}_3
-
(1-\cos(\delta\theta))
{\bf a}_3.
\]

In general, for the $i$th side of the triangle the variation in ${\bf V}$ is
\[
{\bf \delta V}
=
-\sin(\delta\theta)
{\bf b}_i
+
j
(1-\cos(\delta\theta))
{\bf a}_i
\]
where $j=\pm 1$ distinguishes between perturbations due to a rotation about an axis at the head or the tail of the vector ${\bf a}_i$. Expanding to second order in $\delta \theta$ gives,
\[
{\bf \delta V}
\approx
-(\delta\theta)
{\bf b}_i
-
j
\frac{(\delta\theta)^2}{2}
{\bf a}_i.
\]

Thus the length of the periodicity vector is given by
\[
|{\bf V}|
=
\sqrt{
|{\bf V}_o|^2
+
{\bf V}_o
\cdot
\delta V
+
|{\bf \delta V}|^2,
}
\]
expanding in powers of $\delta \theta$ yields  
\[
|{\bf V}|=V_0 + V_1 \delta \theta + V_2 \delta \theta^2É
\]
where the leading order term is
\[
V_0=\sqrt{l^2+n^2-ln}.
\]

The coefficient of the linear term is found to be,
\[
V_1=-\frac{\sqrt{3}}{2}\frac{X_{[i+j]}}{V_0}c_i,
\]
where $c_i$ denote the phyllotactic coefficients as described in section II. The function $X_{[i]}$ is defined as $X_{[1]}=X_{[2]}=1$ and $X_{[3]}=-1$. 

Here, and in what follows, we use square brackets to indicate a modulo function so 
\[
[i]=i \;\textrm{mod}\; 3
\]
such that $[4]=1$ and $[0]=3$. 

The coefficient of the second order term is
\[
V_2
=
\frac{1}{2V_0}
-
\frac{j}{2V_0}
\left( c_{[i+1]} + \frac{1}{2}c_{i} \right)X_{[i+j]}
-
\frac{1}{V_0^3}
\frac{3}{8}
c_{i}^2.
\]

Similarly the average surface density of a triangular lattice is given by
\[
\Pi
=
\left(
\frac{c_{[i+j]}}{\sin(\pi/3)}+\frac{1}{\sin{(\pi/3 + \delta \theta)}}
\right)
\frac{1}{|c_{[i+j]}|}
\]
which can be expanded in terms of $\delta \theta$, so that to second order we have
\[
\Pi
\approx
\Pi_0
+
\Pi_1
\delta \theta
+
\Pi_2
\delta \theta^2
\]
where
\[
\Pi_0=\frac{1}{\sin(\pi/3)}
\]
and
\[
\Pi_1=-\frac{2}{3|c_{[i+j]}|}
\]
and
\[
\Pi_2=\frac{5}{3\sqrt{3}}\frac{1}{|c_{[i+j]}|}.
\]

\subsection*{First order derivatives}

Thus we have
\begin{equation}
\frac{d\Pi}{dV}
=
\frac{\Pi_1}{V_1}
=
\frac{4}{3\sqrt{3}}
\frac{V_0}{|c_{[i+j]}|c_i}
X_{[i+j]}
\label{eq:linear_term}
\end{equation}
The implications of Eq. ($\!\!$~\ref{eq:linear_term}) are as follows. If we confine attention to the linear variation of $\Pi$ with $|{\bf V}|$ around a symmetric structure, there are six possibilities for line-slip, as already stated, which fall into three degenerate pairs. In general two are in the ``forward'' direction, four in the ``backward'' direction, or vice versa. This pattern was evident in our earlier numerical work, but not understood. As for the magnitudes of the slopes, these are given by Eq. ($\!\!$~\ref{eq:linear_term}).

\begin{figure}
\begin{center}
\includegraphics[width=1.0\columnwidth ]{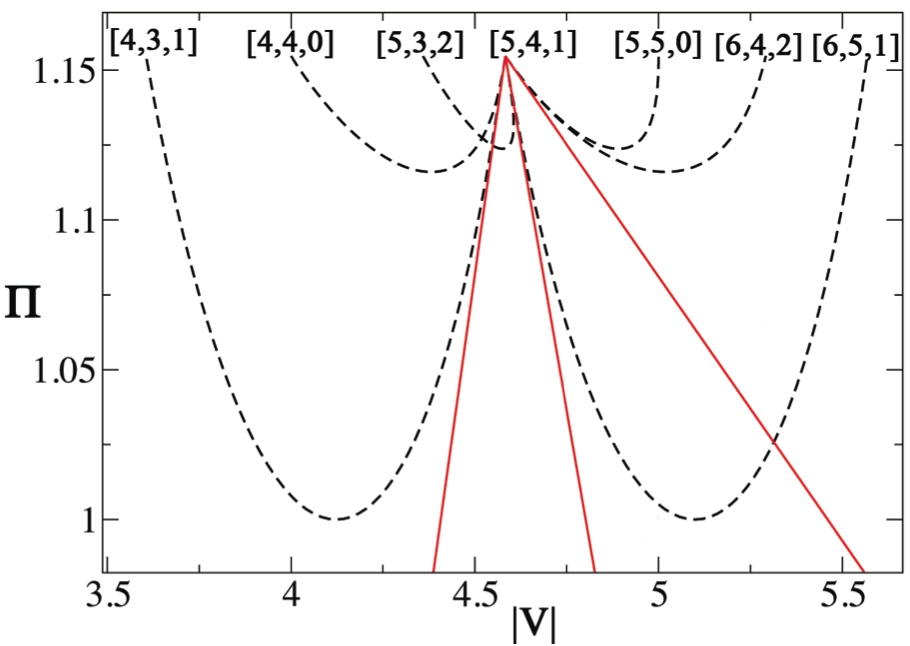}
\caption{A plot of line-slip densities (black dotted curves) against the first order approximations (red lines) close to the symmetric packing [5,4,1].}
\label{line_slip_detail}
\end{center}
\end{figure}

As an example we plot in Fig (\ref{line_slip_detail}) the expansion given by Eq. ($\!\!$~\ref{eq:density_expansion}) up to  the linear term close to the symmetric packing [5,4,1] (see appendix A for the numerical values of the gradients). The expected the gradients are degenerate, so that although in general we expect six distinct line-slip structures there are only three distinct values for the gradients.\\     

\subsection*{Second order derivatives}

Clearly the gradient of the density of the various line-slip packings does not fully determine which line-slip packing has the highest density close to the maximal packing point. In order to distinguish between these proceed to the second derivative, which is evaluated as,
\begin{widetext}
\begin{equation}
\frac{d^2\Pi}{dV^2}
=
\frac
{4X_{[i+j]}(3c_i^2+2c_i(5+j)V_0^2X_{[i+j]} +4V_0^2(jc_{[i+1]}X_{[i+j]} -1))}
{9\sqrt{3}c_i^3|c_{[i+j]}|}
\label{eq:curvature_eq}
\end{equation}
\end{widetext}

\begin{figure}
\includegraphics[width=1.0\columnwidth ]{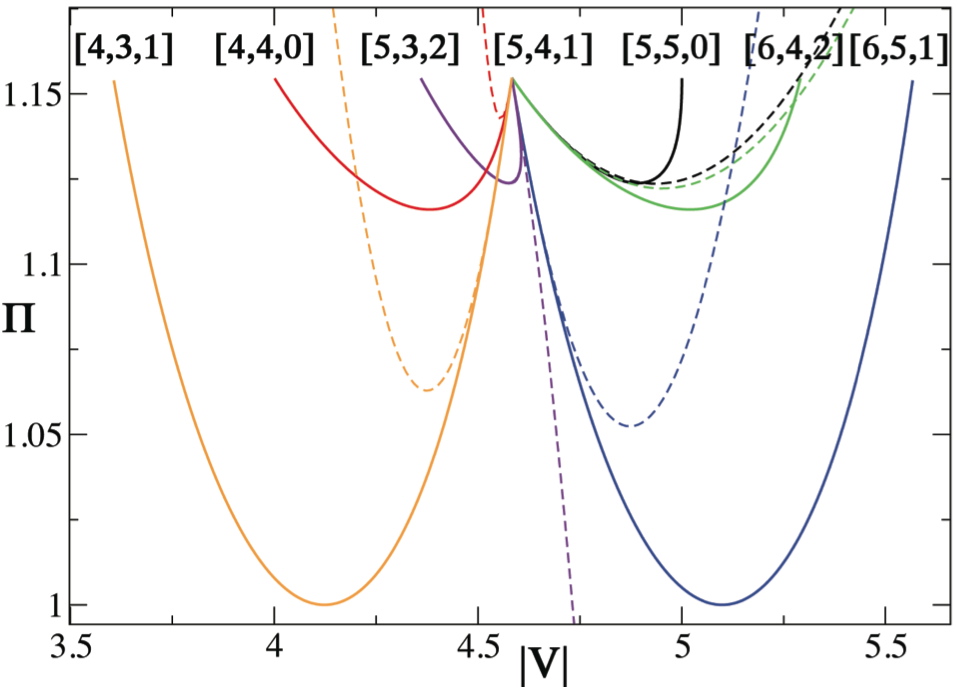}
\caption{Expansion up to second order for the density of line-slip packings close to the symmetric structure [5,4,1]. The solid lines show the densities of the six line-slip structures while the corresponding dashed line of the same colour shows the second order expansion given by Eq. ($\!\!$~\ref{eq:density_expansion})  }
\label{curvature_expansion}
\end{figure}

In Fig (\ref{curvature_expansion}) we plot the expansion given Eq. ($\!\!$~\ref{eq:density_expansion}) up to the second order term (again see appendix A for the numerical values) close to the symmetric packing [5,4,1]. Each line-slip packing is shown by a solid coloured curve and the corresponding approximation is give by a dashed line of the same colour. 

Thus the expansion Eq. ($\!\!$~\ref{eq:density_expansion}) can be used to interpolate between the symmetric close-packed structures. This procedure can reproduce very well the extensive results previously reported for the densities of the intermediate line-slip structures.

\section{Conclusions}

The problem of disk packing on a cylinder has turned out to be surprisingly rich in detail. In this paper we have shown that much of this can be accounted for analytically, offering definite rules for the densest structures. 

We were originally led into this subject by a study of sphere packings in cylinders. The present results also help to shed some light on the corresponding sphere packings, at least qualitatively.

When we began the analysis it seemed only relevant to cylindrical packings of hard disks or spheres. That is the occurrence of the line-slip structures was seen as a feature to be associated with perfectly hard constituents. This is not quite correct. Systems comprised of softly interacting particles, such as those studied by Wood, et al \cite{wood2013self}, can exhibit this feature and we will pursue this in future work using the present results as a starting point.

The study of soft spheres may be a useful approach to relate the present study to related systems in cylindrical confinement such as dry foams \cite{Pittet:1996}, \cite{Weaire:1992}, \cite{Boltenhagen:1998} and wet foams \cite{Mughal:2012}.

\section{Acknowledgements}
AM acknowledges the support of the German Science Foundation (DFG) through the research group ``Geometry and Physics of Spatial Random Systems'' under grant no SCHR-1148/3-1. DW acknowledges the hospitality of Institut f\"{u}r Theoretische Physik I, Fried.-Alex.-Universit\"{a}t Erlangen-N\"{u}rnberg.

\section*{Appendix A}

Taking for example the symmetric packing $[5,4,1]$ we evaluate the first and second derivatives (as described in section V). The table below gives the numerical values of the derivatives, accompanied by the appropriate indices as used in Eq. ($\!\!$~\ref{eq:linear_term}) and Eq. ($\!\!$~\ref{eq:curvature_eq}).
\\

\begin{tabular}{ | l | c | c | c | c | c | c | c | c | r } \hline
  $(l',m',n')$ & $i$ & $j$ & $c_i$  & $c_{[i+1]}$ & $c_{[i+j]}$ & $X_{[i+j]}$ & $ \frac{d\Pi}{dV} $ & $ \frac{d^2\Pi}{dV^2}$\\ \hline
  $[4,4,0]$ & 1 & 1 & 1 & 4 & 4& 1 & 0.882 & 32.524\\
  $[5,3,2]$ & 1 & -1 & 1 & 4 &-5 & -1 & -0.706 & -4.465\\ \hline
  $[5,5,0]$ & 2 & 1 & 4 & -5 & -5& -1 & -0.176& 0.500\\
  $[4,3,1]$ & 2 & -1 & 4 & -5 & 1& 1 & 0.881& 4.234\\ \hline
  $[6,5,1]$ & 3 & 1 & -5 & 1 & 1& 1 & -0.706& 2.433\\
  $[6,4,2]$ & 3 & -1 & -5 & 1 & 4& 1 & -0.176& 0.479\\ \hline 
\end{tabular}

\newpage
\bibliographystyle{nonspacebib}
%\bibliography{mybibliography}

\end{document}